\documentclass[conference,10pt,letterpaper]{IEEEtran}
\usepackage[T1]{fontenc}
\usepackage{cite}
\usepackage{amsmath,amssymb}
\usepackage{graphicx}
\usepackage{booktabs}
\usepackage[hidelinks]{hyperref}
\begin{document}
\title{MirrorComm: A Communication-Efficient Cooperative Perception Framework via Mirrored World Models}
\author{\IEEEauthorblockN{Shenghan Luo\textsuperscript{1}, Yin Xu\textsuperscript{1}, Cixiao Zhang\textsuperscript{1}, Yang Wang\textsuperscript{1}, Yujie Cao\textsuperscript{1}, Zhuoran Xiao\textsuperscript{2}, Wenjun Zhang\textsuperscript{1}}
\IEEEauthorblockA{\textsuperscript{1}Cooperative Medianet Innovation Center (CMIC), Shanghai Jiao Tong University, Shanghai, China\\
\textsuperscript{2}Unisoc (Shanghai) Technologies Co., Ltd., Shanghai, China\\
Corresponding author: Yin Xu (e-mail: xuyin@sjtu.edu.cn)}}
\maketitle
\begin{abstract}
Cooperative perception (CP) requires substantial communication bandwidth to exchange sensing information among connected agents.
A variety of compression techniques have been proposed to reduce transmission costs while retaining task-relevant information.
However, most existing methods follow a frame-wise compression paradigm, overlooking temporal predictability and repeatedly transmitting information that could be inferred from historical context.
To address these limitations, we propose MirrorComm, the first world model-based communication-efficient CP framework.
Specifically, mirrored world models at the transmitter and receiver predict the transmitter's current observation from a shared history of its reconstructed observations.
The transmitter compares this prediction with its current observation and selectively transmits observation updates to correct prediction errors, reducing repeated transmission of predictable information.
Experiments on the OPV2V dataset across two downstream tasks including object detection and semantic segmentation show that MirrorComm reduces communication bandwidth requirements by up to approximately $66\times$ while maintaining comparable performance to a state-of-the-art (SOTA) method.
Ablation studies further demonstrate the effectiveness of the individual components.

\end{abstract}
\begin{IEEEkeywords}
Cooperative perception, communication compression, world models, object detection, semantic segmentation.
\end{IEEEkeywords}
\section{Introduction}
\label{sec:introduction}
Accurate and reliable environmental perception is essential for autonomous agents to make informed decisions.
However, single-agent perception is constrained by limited sensing range and occlusions.
Cooperative perception (CP) mitigates these limitations by enabling agents to exchange sensing information over wireless channels.
This paradigm has been widely applied in vehicular networks, where shared observations reveal objects and regions hidden from an agent's own sensors or beyond its sensing range, enabling more complete and reliable environmental perception~\cite{opv2v,coped,v2vnet}.

However, exchanging sensing information places substantial demands on limited communication bandwidth.
Depending on the information exchanged, CP architectures can be categorized into raw-level, object-level, and feature-level approaches~\cite{taskorientedcp}.
Raw-level methods, such as Cooper~\cite{cooper}, preserve detailed sensor measurements but incur high communication costs.
Object-level methods, such as OptiMatch~\cite{optimatch}, substantially reduce these costs by transmitting compact detection results, but their performance remains constrained by individual agents' local detection capabilities.
Feature-level methods, including F-Cooper~\cite{fcooper}, V2VNet~\cite{v2vnet}, and V2X-ViT~\cite{v2xvit}, balance these considerations by exchanging intermediate representations.
More recent methods, including DiffCP~\cite{diffcp}, ERMVP~\cite{ermvp}, Where2comm~\cite{where2comm}, and CoST~\cite{cost}, further reduce bandwidth requirements through compact feature reconstruction or selective transmission.
However, most existing compression methods follow a frame-wise paradigm, repeatedly transmitting predictable information that could be reconstructed from historical context.

Recent advances in world models offer a promising avenue for exploiting this temporal predictability.
By maintaining internal environmental representations and learning state transitions, world models predict subsequent scene states from historical observations~\cite{baldassarre2025dinoworld,assran2025vjepa2,hafner2025dreamerv3}.
World4V2X~\cite{world4v2x} demonstrates their potential for CP through scene evolution modeling and consistency-guided reconstruction, improving robustness to occlusions and degraded observations.
World models have also been explored in semantic communication to predict video frames at the receiver and reduce frame-by-frame transmission~\cite{jiang2026semanticworldmodels}.
This framework uses textual prompts to guide prediction and relies on depth-map feedback to determine when updates are needed, requiring a feedback link for transmission decisions.
For communication-efficient CP, a key challenge is to enable the transmitter to identify receiver-side prediction errors and selectively transmit observation updates without prediction-quality feedback.

To address this challenge, we propose MirrorComm, a communication-efficient CP framework based on mirrored world models.
At the ego receiving agent, a world model predicts the collaborative transmitting agent's current observation from reconstructed historical observations.
The model combines spatial context at multiple scales with causal temporal attention to capture both local details and scene dynamics for prediction.
The transmitter maintains an identical model and synchronized history, allowing it to reproduce the receiver's prediction and identify discrepancies from its current observation without receiver feedback.
It selectively transmits RVQ-encoded observation features within the communication budget, which both ends use to update the corresponding spatial locations.
The receiver then fuses the reconstructed collaborative observation with its own observation for downstream perception.
This design not only avoids repeated transmission of predictable information but also allows agents to continue predicting scene evolution during communication interruptions, mitigating perception degradation caused by outdated collaborative information.

We summarize the contributions of this work as follows:
\begin{itemize}
\item We propose MirrorComm, to our knowledge, the first world model-based communication-efficient CP framework, which replaces part of the communication with prediction to reduce repeated transmissions. Moreover, it can be integrated into any existing mainstream collaborative algorithm to support diverse downstream tasks.
\item We develop a world model that predicts current observations from historical context through multiscale causal spatiotemporal modeling. To preserve reconstruction fidelity with smaller codebooks, we encode the required observation updates using residual vector quantization (RVQ), which combines codewords across stages.
\item Experiments on OPV2V for object detection and semantic segmentation show that MirrorComm reduces communication bandwidth requirements by up to approximately $66\times$ while maintaining comparable performance to a state-of-the-art (SOTA) method. Additional experiments demonstrate its advantage over the frame-wise paradigm under intermittent communication. Ablation studies further confirm the effectiveness of the proposed modules.
\end{itemize}

\section{Problem Formulation}
\label{sec:problem}
We consider a CP system with $N$ agents that exchange information to support individual local perception.
Each agent can act as a transmitter when sharing information and as a receiver when incorporating messages from its collaborators.
For agent $i$, let $X_i$ denote its current sensory observation (e.g., from LiDAR or cameras) and $Y_i$ denote the ground-truth annotations used to evaluate its perception output.
We use $M_{j\to i}$ to denote the message sent from agent $j$ to agent $i$.
Our goal is to maximize aggregate perception performance under a total communication budget $C$.
Specifically, the optimization problem can be formulated as follows:
\begin{equation}
\begin{aligned}
\max_{\theta,\{M_{j\to i}\}}\quad
& \sum_{i=1}^{N} h\!\left(
\Phi_\theta\!\left(X_i,\{M_{j\to i}\}_{j\ne i}\right),Y_i
\right), \\
\text{s.t.}\quad
& \sum_{i=1}^{N}\sum_{\substack{j=1\\j\ne i}}^{N} c(M_{j\to i})\le C,
\end{aligned}
\label{eq:cp_objective}
\end{equation}
where $\Phi_\theta$ is the perception model with parameters $\theta$, $h(\cdot,\cdot)$ measures task performance, and $c(\cdot)$ quantifies the communication cost of the messages.
The challenge is to design messages $M_{j\to i}$ that both preserve task-relevant information and satisfy the communication budget.

\section{Methodology}
\label{sec:methodology}
\subsection{Overview of MirrorComm Architecture}
\label{sec:method_overview}
\begin{figure*}[!t]
\centering
\includegraphics[width=\textwidth]{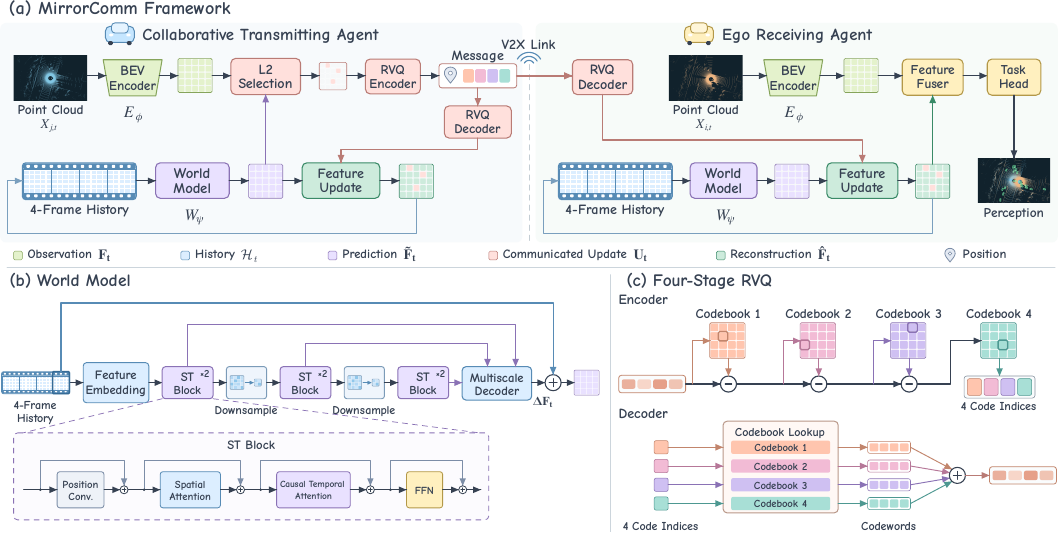}
\caption{Overview of MirrorComm, showing (a) the overall framework, (b) the world model architecture, and (c) four-stage RVQ-based message encoding.}
\label{fig:mirrorcomm_architecture}
\end{figure*}
MirrorComm reduces communication by allowing the receiver to predict the transmitter's current observation and receive selected updates to correct prediction errors.
As illustrated in Fig.~\ref{fig:mirrorcomm_architecture}, each agent first encodes its observation into bird's-eye view (BEV) features, and the receiver predicts the transmitter's current features from four previous reconstructions.
The transmitter maintains an identical world model and synchronized history, enabling it to reproduce this prediction and compare it with its current observation features.
Within the communication budget, it selects the spatial locations with the largest discrepancies and transmits their RVQ-encoded features together with their positions.
Both ends perform the same feature update by replacing the predictions at the selected positions with the decoded observation features and retaining the predictions elsewhere.
Only these reconstructed transmitter features enter the shared histories for subsequent predictions.
The receiver fuses the updated features with its own current features for downstream perception.

By reconstructing predictable information locally, MirrorComm avoids repeatedly transmitting it and concentrates the available bandwidth on observations that are not accurately predicted.
The same predictive mechanism also allows the receiver to continue updating collaborative features during temporary communication outages, mitigating perception degradation caused by outdated collaborative information.
Moreover, MirrorComm naturally supports multi-agent collaboration, with each agent reusing the same world model for receiver-side prediction and transmitter-side replication while maintaining a separate historical context for each communication link.
We next describe MirrorComm using vehicular CP with LiDAR point clouds as an example.

\subsection{BEV Feature Encoder}
\label{sec:bev_encoder}
As a widely adopted representation for CP, BEV preserves the spatial structure of observations and facilitates feature alignment and fusion across agents.
MirrorComm reconstructs the transmitter's observations in this feature space, allowing the reconstructed features to be used by existing BEV-based CP fusion modules.
Specifically, the LiDAR point cloud $X_{i,t}$ observed by agent $i$ at time $t$ is aligned to the receiver's coordinate frame using relative poses, cropped to the region of interest, and discretized into a spatial grid.
The encoder aggregates local point geometry within each grid cell and applies convolutional layers to extract spatial context, yielding the mapping from point clouds to BEV features
\begin{equation}
\mathbf{F}_{i,t}=E_{\phi}(X_{i,t})\in\mathbb{R}^{H\times W\times D},
\label{eq:bev_encoding}
\end{equation}
where $E_{\phi}$ denotes the encoder with parameters $\phi$, including preprocessing, and $H$, $W$, and $D$ denote the BEV feature height, width, and channel dimension, respectively.

\subsection{World-Model-Based Predictive Communication}
\label{sec:predictive_communication}
\subsubsection{Mirrored World Model}
\label{sec:mirrored_world_model}
A world model learns how an environment evolves over time, enabling it to predict future states from historical observations.
In MirrorComm, we use this capability to predict the transmitter's current BEV features from the reconstructed features of the preceding four frames, expressed as
\begin{equation}
\widetilde{\mathbf F}_t=W_{\psi}(\mathcal H_t),
\label{eq:world_model_mapping}
\end{equation}
where $W_{\psi}$ denotes the world model with learnable parameters $\psi$, $\mathcal H_t$ is the shared four-frame historical context, and $\widetilde{\mathbf F}_t$ denotes the predicted BEV features at time $t$.
To capture spatial relationships and temporal evolution jointly, we implement $W_{\psi}$ using a multiscale causal spatiotemporal attention mechanism, as illustrated in Fig.~\ref{fig:mirrorcomm_architecture}(b).

For a single communication link, we omit agent indices for brevity and denote observed and reconstructed BEV features by $\mathbf F_t$ and $\widehat{\mathbf F}_t$, respectively.
To compensate for changes in the reference frame over time, we apply pose-based alignment $\mathcal A_{\tau\rightarrow t}$ to each historical reconstruction $\widehat{\mathbf F}_\tau$, where $\tau\in\{t-4,\ldots,t-1\}$, bringing all four frames into the current coordinate system.
The aligned features, together with their spatial validity masks and elapsed times, form the historical context $\mathcal H_t$ in chronological order.
The masks indicate where the aligned historical features remain spatially valid, while the elapsed times specify the interval between each historical frame and the current frame.

The predictor first embeds the aligned BEV features of each historical frame together with its validity mask and elapsed time into a hidden representation.
The resulting four-frame sequence then passes through three successive stages, each containing two spatiotemporal (ST) blocks.
Between stages, the spatial resolution is halved while all four time steps are retained, yielding outputs $\mathbf Z^{(1)}$, $\mathbf Z^{(2)}$, and $\mathbf Z^{(3)}$ at resolutions $H\times W$, $H/2\times W/2$, and $H/4\times W/4$, respectively.
Fine-resolution features preserve local detail, while coarser features provide broader spatial context for predicting scene evolution.

Within each ST block, spatial and temporal dependencies are modeled sequentially through spatial attention (SA) and causal temporal attention (TA), followed by a feed-forward network (FFN).
For an input feature sequence $\mathbf Z_{\mathrm{in}}$, positional information is first introduced through a depthwise convolution, and the subsequent transformations are applied with separate residual connections as
\begin{equation}
\begin{aligned}
\mathbf Z_{\mathrm{PE}}&=\mathbf Z_{\mathrm{in}}+\operatorname{PE}(\mathbf Z_{\mathrm{in}}),\\
\mathbf Z_{\mathrm{SA}}&=\mathbf Z_{\mathrm{PE}}+\operatorname{SA}(\operatorname{LN}(\mathbf Z_{\mathrm{PE}})),\\
\mathbf Z_{\mathrm{TA}}&=\mathbf Z_{\mathrm{SA}}+\operatorname{TA}(\operatorname{LN}(\mathbf Z_{\mathrm{SA}});\mathbf B),\\
\mathbf Z_{\mathrm{out}}&=\mathbf Z_{\mathrm{TA}}+\operatorname{FFN}(\operatorname{LN}(\mathbf Z_{\mathrm{TA}})),
\end{aligned}
\label{eq:st_block}
\end{equation}
where PE denotes the depthwise positional convolution, LN denotes layer normalization, $\mathbf Z_{\mathrm{out}}$ is the block output, and subscripts $\mathrm{PE}$, $\mathrm{SA}$, and $\mathrm{TA}$ identify the corresponding intermediate stages.
SA aggregates spatial information within each historical frame, using local windows at the two finer scales and global attention at the coarsest scale.
TA then processes the spatially updated sequence along the time dimension at each aligned spatial location, with a causal mask $\mathbf B$ added to the attention scores before softmax, defined by
\begin{equation}
B_{ab}=\begin{cases}
0,&b\leq a,\\
-\infty,&b>a,
\end{cases}
\qquad a,b\in\{1,2,3,4\},
\label{eq:causal_mask}
\end{equation}
where $a$ and $b$ index the query and key time steps in chronological order, respectively, so that each historical step attends only to itself and earlier steps.

After spatiotemporal modeling, the decoder $f_{\mathrm d}$ takes the last temporal output from each scale, denoted $\mathbf Z^{(1)}_4$, $\mathbf Z^{(2)}_4$, and $\mathbf Z^{(3)}_4$, where the subscript $4$ corresponds to time $t-1$.
It progressively upsamples coarser features and combines them with finer features through skip connections to predict a feature residual $\Delta\mathbf F_t$.
Adding this residual to the aligned reconstruction of the preceding frame produces the current prediction, expressed as
\begin{equation}
\begin{aligned}
\Delta\mathbf F_t&=f_{\mathrm d}(\mathbf Z^{(1)}_4,\mathbf Z^{(2)}_4,\mathbf Z^{(3)}_4),\\
\widetilde{\mathbf F}_t&=\mathcal A_{t-1\rightarrow t}(\widehat{\mathbf F}_{t-1})+\Delta\mathbf F_t.
\end{aligned}
\label{eq:world_prediction}
\end{equation}
Together, the history embedding, three ST stages, and residual decoding constitute $W_{\psi}$, which predicts scene changes relative to the preceding reconstruction.

\subsubsection{Prediction-Guided Communication and Feature Update}
\label{sec:prediction_guided_update}
During initialization, the transmitter sends RVQ-encoded features at all spatial locations for four consecutive frames, after which both ends begin prediction from this shared history.

To correct prediction errors within the communication budget, the transmitter compares the predicted features with those extracted from its current observation and assigns each spatial position $p$ an L2 discrepancy score,
\begin{equation}
\epsilon_t(p)=\frac{1}{D}\left\|\mathbf F_t[p]-\widetilde{\mathbf F}_t[p]\right\|_2^2,
\label{eq:update_score}
\end{equation}
where $[p]$ extracts the feature vector at position $p$ and $\|\cdot\|_2$ denotes the Euclidean norm.
After accounting for packet metadata, the transmitter selects the positions with the largest positive scores that fit the link budget, forming $\mathcal S_t$, and transmits their positions and RVQ-encoded observation features as described in Section~\ref{sec:rvq_encoding}.
The budgets allocated to individual links sum to at most the total budget $C$.
Both ends decode these features into $\mathbf U_t\in\mathbb R^{H\times W\times D}$, which contains the recovered features at $\mathcal S_t$ and zeros elsewhere.
Feature Update replaces predictions at selected positions with recovered features and retains predictions elsewhere, yielding
\begin{equation}
\widehat{\mathbf F}_t=\mathbf U_t+(1-\mathbf m_t)\odot\widetilde{\mathbf F}_t,
\label{eq:feature_update}
\end{equation}
where $\mathbf m_t\in\{0,1\}^{H\times W}$ marks the selected positions and $\odot$ denotes element-wise multiplication with channel broadcasting.
This deterministic operation introduces no learnable parameters.
By applying the same feature update, both ends obtain identical reconstructed features $\widehat{\mathbf F}_t$, which are appended to their shared histories while the oldest frame is discarded, preserving a common four-frame context for subsequent predictions.
When no update is transmitted, both ends use $\widehat{\mathbf F}_t=\widetilde{\mathbf F}_t$ to advance their histories through prediction alone.
The receiver fuses $\widehat{\mathbf F}_t$ with its current local features for downstream perception, while retaining only the reconstructed transmitter features in the shared history.

\subsubsection{Closed-Loop Training}
\label{sec:closed_loop_training}
We first train the perception backbone, then train the RVQ codec and adapt the receiver to quantized features as described in Section~\ref{sec:rvq_encoding}.
With the encoder and codebooks frozen, we jointly optimize the predictor, fusion module, and task head by unrolling the prediction--communication--update loop.
Reconstructed features are fed back as history, exposing the predictor to errors accumulated during deployment.
The feature-learning objective $\mathcal L_{\mathrm{WM}}$ combines three losses,
\begin{equation}
\mathcal L_{\mathrm{WM}}=
\lambda_{\mathrm p}\mathcal L_{\mathrm{pred}}
+\lambda_{\mathrm r}\mathcal L_{\mathrm{rec}}
+\lambda_{\mathrm u}\mathcal L_{\mathrm{unrecv}},
\label{eq:world_training}
\end{equation}
where $\lambda_{\mathrm p}$, $\lambda_{\mathrm r}$, and $\lambda_{\mathrm u}$ are nonnegative weights.
The three terms use the current observation features $\mathbf F_t$ as supervision and are defined as
\begin{equation}
\begin{aligned}
\mathcal L_{\mathrm{pred}}&=\operatorname{MSE}(\widetilde{\mathbf F}_t,\mathbf F_t),\\
\mathcal L_{\mathrm{rec}}&=\operatorname{SmoothL1}(\widehat{\mathbf F}_t,\mathbf F_t),\\
\mathcal L_{\mathrm{unrecv}}&=\operatorname{MSE}_{1-\mathbf m_t}(\widehat{\mathbf F}_t,\mathbf F_t),
\end{aligned}
\label{eq:reconstruction_losses}
\end{equation}
where $\operatorname{MSE}$ and $\operatorname{SmoothL1}$ denote the mean squared error and smooth $L_1$ loss averaged over feature elements, respectively.
The latter uses a transition threshold of one.
The mask subscript restricts the average to untransmitted positions, with the loss set to zero when none remain.
These terms supervise prediction before communication, reconstruction after feature updates, and prediction at untransmitted positions, respectively.
The joint objective adds the downstream task loss on the final perception output to $\mathcal L_{\mathrm{WM}}$.
The feature-learning term supervises the predictor, while the task loss trains both the predictor and receiver through reconstructed features.

\begin{figure*}[!t]
\centering
\includegraphics[width=\textwidth]{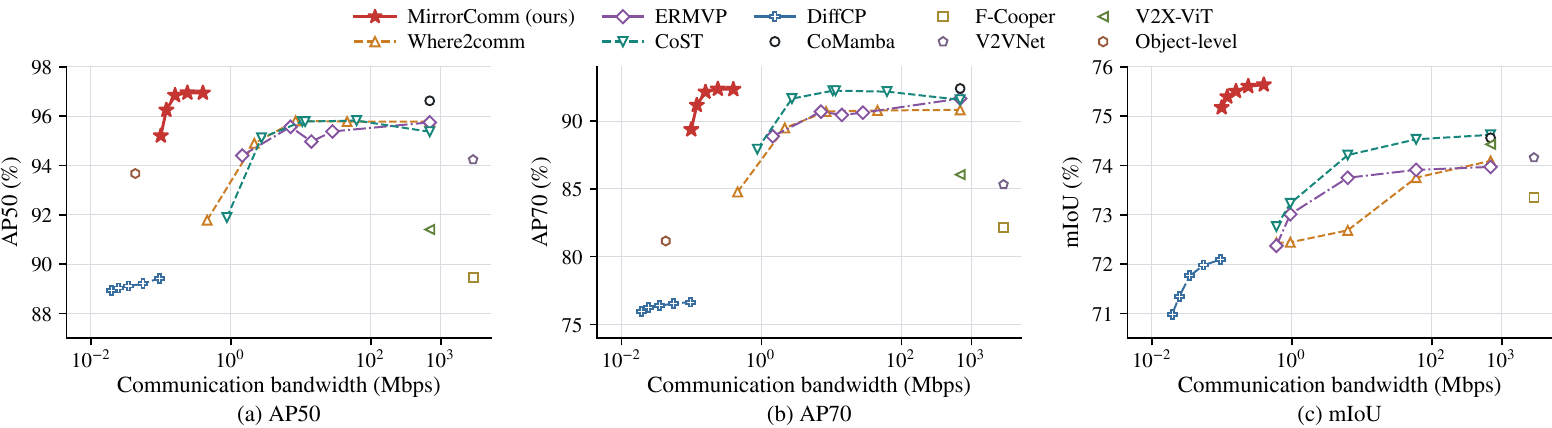}
\caption{Trade-off between perception performance and communication bandwidth on OPV2V. Panels (a) and (b) show object detection AP at IoU thresholds of 0.5 and 0.7, respectively, while panel (c) shows mIoU for BEV semantic segmentation.}
\label{fig:main_tradeoff}
\end{figure*}

\subsection{RVQ-Based Message Encoding}
\label{sec:rvq_encoding}
To communicate observation updates efficiently, we seek a compact encoding that preserves reconstruction fidelity with smaller codebooks.
MirrorComm therefore uses four-stage RVQ, which combines codewords across stages by successively quantizing the remaining residual and transmits their indices (Fig.~\ref{fig:mirrorcomm_architecture}(c)).
Each stage $\ell\in\{1,2,3,4\}$ uses a codebook $\mathcal C_\ell=\{\mathbf c_{\ell,k}\}_{k=1}^{K}\subset\mathbb R^D$, where $K$ is the number of codewords and $\mathbf c_{\ell,k}$ is the $k$th codeword.
The four codebooks provide up to $K^4$ index combinations with only $4K$ stored codewords.

For each selected position $p\in\mathcal S_t$, the transmitter initializes $\mathbf r_0=\mathbf F_t[p]$ and recursively quantizes the residual as
\begin{equation}
\begin{aligned}
k_\ell&=\operatorname*{arg\,min}_{1\le k\le K}\|\mathbf r_{\ell-1}-\mathbf c_{\ell,k}\|_2^2,\\
\mathbf r_\ell&=\mathbf r_{\ell-1}-\mathbf c_{\ell,k_\ell},\qquad \ell=1,2,3,4,
\end{aligned}
\label{eq:rvq_encoding}
\end{equation}
where $k_\ell$ is the selected index and $\mathbf r_\ell$ the remaining residual, with time and position indices omitted for brevity.
The transmitter sends the four codeword indices, which both ends decode using identical preinstalled codebooks as
\begin{equation}
\mathbf U_t[p]=\sum_{\ell=1}^{4}\mathbf c_{\ell,k_\ell},\qquad p\in\mathcal S_t.
\label{eq:rvq_reconstruction}
\end{equation}
Unselected positions are set to zero to form $\mathbf U_t$ for the feature update in~\eqref{eq:feature_update}.
For a link budget $C_{\mathrm{link}}$, the packet satisfies $b_{\mathrm h}+b_{\mathrm r}|\mathcal S_t|\le C_{\mathrm{link}}$, where $b_{\mathrm h}$ and $b_{\mathrm r}$ denote the metadata bytes and bytes per selected position, respectively.

The codec is trained using standard codebook-fitting and commitment losses~\cite{oord2017vqvae}, together with feature reconstruction and downstream task losses.
Codewords are detached when forming subsequent residuals, while a straight-through estimator propagates gradients through quantization to the encoder.

\section{Experimental Results}
\label{sec:experiments}
\subsection{Experimental Setup}
\label{sec:experimental_setup}
We conduct experiments on OPV2V~\cite{opv2v}, a simulated CP dataset generated with CARLA and OpenCDA.
Our processed splits contain 6,374 training frames, 1,980 validation frames, and 2,170 test frames.
We use its LiDAR sequences with a perception frequency of $f=10$~Hz and process frames within each scene in chronological order.
The shared histories are reset when switching to a new scene.
The communication bandwidth is measured in Mbps as
\begin{equation}
B_{\mathrm{comm}}=\frac{8fB_{\mathrm{total}}}{10^6N_{\mathrm f}},
\label{eq:communication_bandwidth}
\end{equation}
where $B_{\mathrm{total}}$ is the total number of transmitted bytes, including metadata, over $N_{\mathrm f}$ evaluated frames.
For MirrorComm, this total also includes the cold-start transmissions used to initialize the shared history at the beginning of each scene.

Two downstream tasks are evaluated on this dataset, including object detection and BEV semantic segmentation.
These tasks assess complementary aspects of reconstruction quality, as object detection requires accurate object locations and geometry, whereas semantic segmentation requires spatially dense semantic details.
For object detection, we report BEV average precision (AP) at intersection-over-union (IoU) thresholds of 0.5 and 0.7, denoted AP50 and AP70, respectively.
For semantic segmentation, we report mean IoU (mIoU) over the vehicle, road, and lane classes, excluding background.

We implement MirrorComm with the feature fusion module from CoMamba~\cite{comamba}.
We evaluate per-link update budgets of 256, 512, 1,024, 2,048, and 4,096 bytes.
We use AdamW with an initial learning rate of $10^{-4}$ and weight decay of $10^{-4}$, and apply truncated backpropagation through time over 20 steps of the closed-loop sequence.
All implementations use PyTorch and run on NVIDIA RTX~4090 GPUs with 24~GB memory.

\subsection{Trade-off between Performance and Bandwidth}
We compare with representative feature-level information fusion methods for CP, including V2VNet~\cite{v2vnet}, F-Cooper~\cite{fcooper}, V2X-ViT~\cite{v2xvit}, and CoMamba~\cite{comamba}, which fuse intermediate features from multiple agents.
We also include methods targeting communication efficiency, namely DiffCP~\cite{diffcp}, ERMVP~\cite{ermvp}, Where2comm~\cite{where2comm}, and CoST~\cite{cost}.
For object detection, we additionally consider an object-level baseline that transmits local detection results without a bandwidth cap.

Fig.~\ref{fig:main_tradeoff} shows that MirrorComm substantially reduces communication bandwidth while preserving perception performance across both downstream tasks.
At an update budget of 1,024 bytes, MirrorComm requires only 0.159~Mbps and achieves 96.84\% AP50 and 92.15\% AP70.
Compared with the state-of-the-art method CoST, which achieves 95.78\% AP50 and 92.24\% AP70 at 10.497~Mbps, MirrorComm delivers higher AP50 and comparable AP70 while requiring approximately $66\times$ less bandwidth.
This advantage extends to semantic segmentation, where MirrorComm achieves 75.51\% mIoU at the same bandwidth, surpassing CoST's best evaluated mIoU of 74.62\% while operating below its lowest evaluated bandwidth of 0.603~Mbps.
MirrorComm also exceeds frame-wise CoMamba's mIoU of 74.56\%, which may stem from its use of historical context to complement current observations.
Although DiffCP operates at lower bandwidths than MirrorComm, its perception performance is substantially lower on both tasks.
Likewise, the object-level baseline requires only 0.043~Mbps but achieves an AP70 of 81.15\%, revealing a substantial accuracy gap under the stricter localization criterion.
Moreover, unlike most existing methods that operate with fixed data rates, MirrorComm can adapt to different communication budgets at inference time by varying the number of spatial positions selected for transmission, without retraining.

\subsection{Ablation Studies}
\label{sec:component_ablation}
We conduct ablation studies by comparing four component configurations at a fixed communication update budget of 1,024 bytes, reporting AP50, AP70, and mIoU in Table~\ref{tab:semantic_ablation}.
The configurations enable or disable RVQ and the world-model predictor (WM).
Disabling RVQ replaces it with single-stage VQ.
Disabling WM replaces prediction with the pose-aligned previous state, while retaining the same feature update at transmitted locations.
Enabling both components gives the complete MirrorComm framework.
Within each task and quantizer pair, both WM settings use the same receiver weights and communication budget.
VQ uses one 1,024-entry codebook, whereas RVQ uses four 256-entry codebooks, keeping total codeword counts equal and accounting for index lengths in the communication cost.

\begin{table}[t]
\centering
\caption{ABLATION STUDY RESULTS OF THE WORLD MODEL AND RVQ.}
\label{tab:semantic_ablation}
\footnotesize
\setlength{\tabcolsep}{9pt}
\begin{tabular}{@{}ccrrr@{}}
\toprule
\multicolumn{2}{c}{Component} & \multicolumn{3}{c}{Performance (\%)} \\
\cmidrule(lr){1-2}\cmidrule(l){3-5}
RVQ & WM & AP50 & AP70 & mIoU \\
\midrule
$\times$ & $\times$ & 95.12 & 87.01 & 71.09 \\
$\times$ & $\checkmark$ & 96.18 & 89.11 & 72.30 \\
$\checkmark$ & $\times$ & 96.19 & 90.12 & 72.43 \\
$\checkmark$ & $\checkmark$ & \textbf{97.14} & \textbf{92.15} & \textbf{75.51} \\
\bottomrule
\end{tabular}
\end{table}

WM improves both downstream tasks with either quantizer, increasing mIoU by 3.08 percentage points with RVQ.
With WM enabled, replacing VQ with RVQ further improves all three metrics, including a 3.21-point mIoU gain.
These results confirm that both components contribute to perception accuracy under the same communication budget.

\subsection{Robustness to Intermittent Communication}
To evaluate MirrorComm's ability to mitigate information staleness relative to frame-wise perception, we test communication updates at fixed intervals of 2--5 frames and random intervals within this range.
We choose CoMamba as the baseline because MirrorComm adopts its feature fusion architecture.
CoMamba transmits complete collaborative features at each update and holds them between updates.
Under the same schedule, MirrorComm sends 1,024-byte observation updates and predicts collaborative features between transmissions.
Both methods are evaluated on object detection using AP70.

\begin{table}[t]
\centering
\caption{AP70 (\%) under intermittent communication.}
\label{tab:intermittent_communication}
\footnotesize
\setlength{\tabcolsep}{8pt}
\begin{tabular}{@{}ccc@{}}
\toprule
Update interval (frames) & CoMamba & MirrorComm \\
\midrule
2 & 83.78 & \textbf{84.79} \\
3 & 80.69 & \textbf{83.93} \\
4 & 79.11 & \textbf{83.12} \\
5 & 78.67 & \textbf{82.17} \\
Random 2--5 & 78.01 & \textbf{83.54} \\
\bottomrule
\end{tabular}
\end{table}

As shown in Table~\ref{tab:intermittent_communication}, MirrorComm outperforms frame-wise CoMamba under every update schedule.
As the fixed interval increases from 2 to 5 frames, its AP70 decreases by 2.62 percentage points, compared with 5.11 points for frame-wise CoMamba.
Under random intervals, MirrorComm achieves 83.54\% AP70, exceeding the baseline by 5.53 percentage points.
These results show that predicting scene evolution mitigates degradation from stale collaborative information, even though frame-wise CoMamba receives complete collaborative features at every update.

\section{Conclusion}
\label{sec:conclusion}
This paper presented MirrorComm, a world-model-based communication-efficient CP framework.
By maintaining mirrored world models at the transmitter and receiver, MirrorComm reconstructs predictable information locally and communicates compact observation updates to correct prediction errors.
Experiments on OPV2V demonstrated substantial bandwidth savings while maintaining competitive object detection and semantic segmentation performance, with additional benefits under intermittent communication.
These findings highlight the potential of shared predictive models to shift cooperative perception from transmitting observations frame by frame toward communicating only the information needed to update a shared reconstruction.

\end{document}